\documentclass[aps,prb,twocolumn,superscriptaddress,showpacs,10pt]{revtex4-1}
\usepackage{graphicx}
\usepackage{hyperref}

\newcommand{\Ei}{E_i}

\begin{document}


\title{
Depth dependence of the ionization energy of shallow hydrogen states in ZnO and CdS
}



\author{T.~Prokscha}
\email{thomas.prokscha@psi.ch}
\affiliation{Laboratory for Muon Spin Spectroscopy, Paul Scherrer Institut, CH-5232 Villigen PSI, Switzerland}
\author{H.~Luetkens}
\affiliation{Laboratory for Muon Spin Spectroscopy, Paul Scherrer Institut, CH-5232 Villigen PSI, Switzerland}
\author{E.~Morenzoni}
\affiliation{Laboratory for Muon Spin Spectroscopy, Paul Scherrer Institut, CH-5232 Villigen PSI, Switzerland}
\author{G.J.~Nieuwenhuys}
\email[Deceased.]{}
\affiliation{Laboratory for Muon Spin Spectroscopy, Paul Scherrer Institut, CH-5232 Villigen PSI, Switzerland}
\affiliation{Kamerlingh Onnes Laboratory, Leiden University, 2300 RA Leiden, The Netherlands}
\author{A.~Suter}
\affiliation{Laboratory for Muon Spin Spectroscopy, Paul Scherrer Institut, CH-5232 Villigen PSI, Switzerland}
\author{M.~D\"obeli}
\affiliation{Ion Beam Physics, ETH Zurich, CH-8093 Zurich, Switzerland}
\author{M.~Horisberger}
\affiliation{Laboratory for Developments and Methods, Paul Scherrer Institut, CH-5232 Villigen PSI, Switzerland}
\author{E.~Pomjakushina}
\affiliation{Laboratory for Developments and Methods, Paul Scherrer Institut, CH-5232 Villigen PSI, Switzerland}

\date{\today}

\begin{abstract}
The characteristics of shallow hydrogen-like muonium (Mu) states in nominally undoped 
ZnO and CdS (0001) crystals have been studied close to the surface at depths in 
the range of 10~nm -- 180~nm by using low-energy muons, and in the bulk using conventional $\mu$SR. 
The muon implantation depths are adjusted by tuning the energy of the low-energy muons between
2.5~keV and 30~keV. We find that 
the bulk ionization energy $\Ei$~of the shallow donor-like Mu state is lowered 
by about 10~meV at a depth of 100~nm, and continuously decreasing on approaching the surface. 
At a depth of about 10~nm $\Ei$~is further reduced by 25 -- 30~meV compared 
to its bulk value. We attribute this change to the presence of electric fields due to band 
bending close to the surface, and we determine the depth profile of the electric field within
a simple one-dimensional model.
\end{abstract}

%
%
%

\maketitle

%
%
\section{Introduction}\label{introduction}
For future semiconductor technologies the incorporation, profiling and monitoring of dopants is considered
to be a key issue for novel device applications \cite{koenraad_single_2011}. The binding energy of a dopant 
is an important characteristic, and recent studies have focused on the investigation of the binding energies
of single hydrogenic defect states close to semiconductor surfaces, in nanoscale devices, or in quantum wells
\cite{sellier_transport_2006,perraud_direct_2008,teichmann_controlled_2008,wijnheijmer_enhanced_2009}. 
For a shallow Coulombic hydrogenic impurity state in the effective mass approximation, the binding
energy of the state is predicted to decrease when approaching a potential barrier at a
semiconductor interface or surface \cite{bastard1988,mailhiot_energy_1982}. A model proposed by 
Levine\cite{levine_nodal_1965} showed that at a semiconductor surface the ground state of a shallow
impurity is the 2p state, which means that the binding energy of the surface donor is 1/4 of the bulk
donor. However, image charges at the surface cause the binding energy to be closer to its
bulk value, while still being reduced \cite{jiang_shallow_1985}. In contrast, recent theoretical studies
found an \textit{increase} of the hydrogenic impurity binding energies in nanowires and quantum dots due
to dielectric confinement \cite{diarra_ionization_2007,peter_effect_2009}. Additionally, an experimental
investigation of Si doped GaAs estimated an increasing binding energy at 
depths $z < 1.5$~nm \cite{wijnheijmer_enhanced_2009}, and the authors concluded that the effective mass approach
will fail for all hydrogenic donors close to a semiconductor surface.

The descibed effects so far occur on a length scale of typically less than ten nanometers. In this paper we will
discuss the change of the binding energy of single shallow hydrogen-like donor states on a much larger length 
scale at 
depths between 10~nm and $\sim$~200~nm, where the effective mass approximation appears to be appropriate, and 
dielectric confinement and position dependent effective masses or dielectric constants are negligible. 
Even in the case of band bending at the surface -- which is of particular importance 
for this work -- the effective mass approximation is expected to hold because the fractional change of the 
perturbation potential over the dimension of a unit cell is negligibly 
small\cite{zak_effective-mass_1966,luttinger_motion_1955} (less than 1\% in our case).
In case of good sample quality with dislocation line densities $\lesssim 10^5$/cm$^2$ the effect of
internal strain, due to these dislocations, on the ionization energy of shallow donor states can be neglected 
\cite{kohn_shallow_1957}. 
However, the donor ionization energy may be affected by the presence of electric fields close to the 
sample surface: in a lowly doped semiconductor, surface states can cause a pinning of the Fermi level at 
the surface which results 
in a band bending on a length scale given by the Debye length \cite{sze_physics_2006}.
In this case the solution of the Poisson equation yields an quadratic $z$ dependence of the 
electrostatic potential
in the band bending zone close to the surface \cite{sze_physics_2006}, which means a linearly increasing 
electric field towards the surface. The presence of an electric field lowers the ionization energy of the
shallow impurity (Poole-Frenkel effect in insulators or 
semiconductors\cite{sze_physics_2006,frenkel_pre-breakdown_1938,hartke_threedimensional_1967,martin_electric_1981}),
and -- since the electric field increases on approaching the surface -- the binding energy is expected to
decrease when getting closer to the surface.

Hydrogen as an ubiquitous impurity is of particular interest in semiconductor technology, 
because it often modifies the electrical and optical
properties in an unwanted way due to its amphoteric behaviour, which may cause doping counteracting
the prevailing type of conductivity. The characterization of hydrogen impurities in semiconductors
is often difficult, particularly if one wants to study single (or solitary) dopants. Here, positively charged muons 
($\mu^+$) can help to circumvent these difficulties. Muons played an important role in the identification and 
characterization of isolated hydrogen defect centers in semiconductors 
\cite{patterson_muonium_1988,chow1998,cox_shallow-to-deep_2003,cox_muonium_2009}. 
Implanted in a semiconductor or insulator the $\mu^+$ stops at an interstitial site, and 
may capture one or two electrons to form the 
light hydrogen pseudo-isotope muonium [Mu, ($\mu^+e^-$), mass of $\mu^+ \simeq 1/9$ proton masses]. 
Depending on the concentration of other dopants, and on Mu formation energy, it occurs 
in either of three charge states Mu$^+$, Mu$^0$ or Mu$^-$, analogous to hydrogen.
The neutral state can be spectroscopically distinguished from the charged states in muon spin rotation
experiments ($\mu$SR) \cite{yaouanc_muon_2011}. The recent theoretical discovery of a universal alignment of the 
so-called hydrogen pinning level $\varepsilon$(+/-) -- where the formation energies of the 
positive and negative impurity are equal -- allows predictions whether hydrogen forms a shallow donor: 
this occurs if the pinning level is close to or above the conduction band minimum 
\cite{kilic_n-type_2002,van_de_walle_universal_2003}.
The predicted shallow hydrogen donor states in ZnO \cite{van_de_walle_hydrogen_2000}
and InN \cite{van_de_walle_universal_2003} have been first confirmed by $\mu$SR measurements 
\cite{cox_experimental_2001,davis_shallow_2003},
closely followed in ZnO by EPR \cite{hofmann_hydrogen:relevant_2002} and infrared spectroscopy
\cite{mccluskey_infrared_2002,lavrov_hydrogen-related_2002}. In CdS the observed shallow Mu state
\cite{gil_novel_1999,gil_shallow_2001} does not necessarily imply, according to the theoretical models,
that in thermodynamic equilibrium hydrogen acts as a shallow donor. Since the muon experiments
take place on a microsecond time scale (muon life time is $\sim$~2.2~$\mu$s), the observed shallow
Mu state could be a metastable state \cite{shimomura_muonium_2004,van_de_walle_universal_2006,lichti_hydrogen_2008}.

\begin{figure}[ht]
\includegraphics[width=0.95\linewidth]{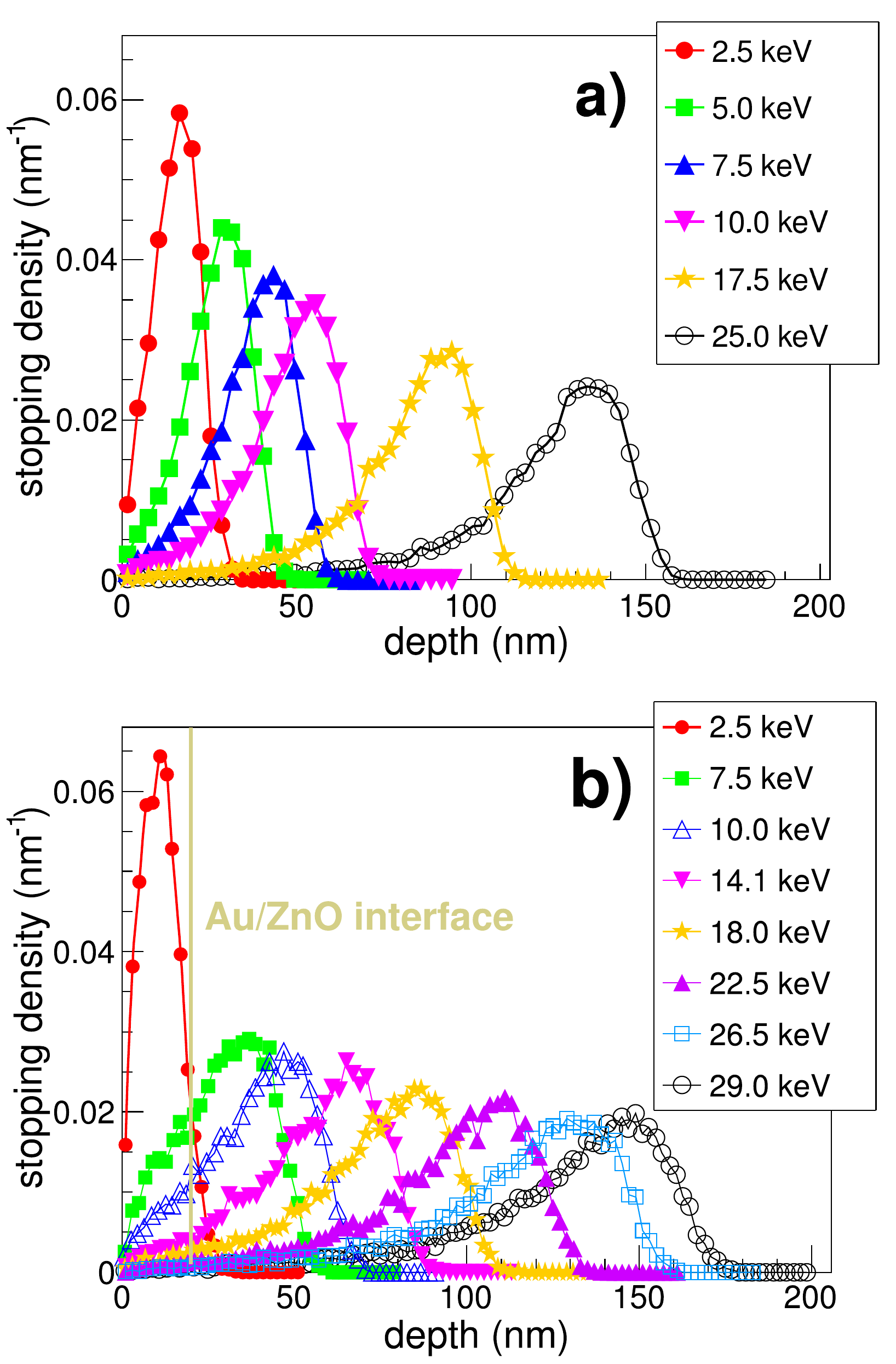}
\caption{Simulated muon implantation profiles in a) ZnO, and b) 20 nm Au on top of
ZnO, using the program TrimSP\cite{trimsp} which has been shown to calculate the stopping
profile with sufficient accuracy\cite{morenzoni_implantation_2002}.}\label{MuonStoppingProfile}
\end{figure}

In this article we present the depth dependence of the ionization energy of the shallow Mu state in CdS 
and of the shallow donor state in ZnO, and at the interface of a 20-nm-thin Au film sputtered on ZnO. This allows us 
to obtain direct information about the effect of electric fields -- due to band bending -- on the ionization 
energies of the corresponding hydrogen-like states in a range of $\sim$~200~nm beneath the surface, 
or at a metal-semiconductor interface. The ionization energies continuously decrease on approaching the 
surface/interface, reaching a reduction of 25 -- 30~meV at a depth
of 10~nm. We use the observed ``ionization profile'' to determine by a simple one-dimensional model the electric 
field profile at the surface/interface. This is to best of our knowledge the first time, that the 
``ionization profile'' 
of a single impurity and the derived electric field profile has been visualized by means of a 
\textit{local probe} implantation technique.
%
%
It offers several advantages compared to 
conventional experimental techniques. Photo-emission spectroscopy and other surface sensitive techniques (see 
the review of Koenraad and Flatt\'{e} ~\cite{koenraad_single_2011} and references therein) are limited to a 
few surface layers, and
cannot access interfaces at a depth of tens of nanometers. Deep-level transient spectroscopy is not applicable
to low-doped materials, shallow impurity states, and single dopants, and usually requires a p-n junction.
In contrast, there are no such limitations for muon spin rotation, which is contactless, and provides intrinsic 
information, about shallow as well as deep hydrogen states. They are incorporated as solitary dopants, and
their properties can be monitored as a function of distance to the surface or an interface, thus providing a
new experimental tool to address the issues raised at the beginning of this introduction.

\section{Experimental details}
%
%
%
The $\mu$SR experiments were carried out at the Swiss Muon Source (S$\mu$S, Paul Scherrer Institut,
Villigen, Switzerland). For the near-surface measurements at mean depths $\langle z\rangle < 200$~nm
we used the low-energy $\mu$SR spectrometer (LE-$\mu$SR) LEM at the $\mu$E4 beam line 
\cite{morenzoni_low-energy_2000,prokscha_new_2008}. Polarized low-energy $\mu^+$ with energies in 
the keV range are generated by moderation of a 4 MeV muon beam in a solid, about 250-nm-thin Ar film with 
a $\sim 10$-nm-thin N$_2$ capping layer, deposited at 10~K on a patterned Ag foil
\cite{harshman_generation_1987,morenzoni_generation_1994,prokscha_moderator_2001,morenzoni_nano-scale_2004}.
Epithermal muons with a mean energy of $\sim 15$~eV escape into vacuum with a conversion efficiency from
MeV-to-eV of $\sim 5\times 10^{-5}$. They are accelerated electrostatically to energies up to 20~keV by
applying a positive bias to the Ag moderator foil, and then transported by use of electrostatic elements 
over a distance of about two meters  to the sample cryostat.
The muon implantation energy was varied between 2.5 keV and 30 keV, corresponding
to mean implantation depths of 10 nm and 150 nm, respectively, see Fig.~\ref{MuonStoppingProfile}.
The implantation energy is usually varied by applying a positive or negative bias of up to 12~kV
to the electrically insulated sample holder \cite{morenzoni_low-energy_2000}.
Shallow Mu formation deep in the bulk at $\langle z\rangle \sim 200-300$~$\mu$m
was studied at the GPS spectrometer at the $\pi$M3 beam line with a muon beam energy of about 3.5~MeV. 

The samples were nominally undoped ZnO and CdS wafers [(0001) orientation, supplier: 
Crystec GmbH, Berlin, Germany; resistivity $> 10$~$\Omega$cm and $> 1$ k$\Omega$cm, respectively]. 
The ZnO crystals had a size of $10\times10$~mm$^2$ and a thickness of 0.5 mm. Nine pieces were glued with
conductive silver onto a standard sample plate of LEM. This
mosaic of samples ensured that no muons missed the sample, therefore
eliminating any background contribution (the low-energy muon beam spot has a FWHM of
about 13 mm). In a 2nd experiment several pieces of the ZnO crystals were sputtered with a 20-nm-thin
Au film to study any changes introduced by the Schottky barrier at the Au/ZnO interface.
The CdS sample was one half of a 2'' wafer, 0.5 mm thick which was
also large enough to stop all muons in the sample. All samples were polished on
both sides. For the GPS measurements one of the ZnO crystals was used, and a $\sim 5\times 10$~mm$^2$
broken-off piece of the CdS wafer.

Transverse field $\mu$SR measurements have been performed with a magnetic field of 10 mT
applied parallel to the $\langle0001\rangle$ direction, and the muon spin initially parallel
to the sample surface at LEM, and out of plane at GPS. Shallow Mu in CdS and ZnO has an anistropic, 
axially symmetric hyperfine interaction
with the hyperfine coupling constants $A_{\parallel}$ and $A_{\perp}$ parallel and perpendicular
to the symmetry axis, which is along the Cd-S or the Zn-O bond direction\cite{gil_novel_1999,gil_shallow_2001}.
In the high field limit two shallow Mu lines can be observed with a separation
$\Delta\nu(\Theta) = A_{\parallel} \cos^2\Theta + A_{\perp} \sin^2 \Theta$, where $\Theta$ is 
the angle between the Mu symmetry axis and the externally applied field. The two lines are placed symmetrically
around the ``central'' line, i.e. the precession frequency of bare $\mu^+$ without bound electron. 
In the geometry of the experiment there is one shallow Mu state at a bond parallel to the 
$\langle0001\rangle$ direction ($\Theta = 0$), and three Mu states at the bonds under 
$\Theta = 109.4^\circ$ with respect to  the $\langle0001\rangle$ axis. This leads to two pairs of 
Mu lines with an intensity ratio of 1:3. For ZnO and CdS the hyperfine
couplings are $A_{\parallel} = 760(30)$~kHz and $335(8)$~kHz,  $A_{\perp} = 370(22)$~kHz and $199(6)$~kHz, 
respectively\cite{gil_shallow_2001}.
However, only in CdS the Mu lines are narrow and well resolved in bulk $\mu$SR
experiments whereas in ZnO, spin- and/or charge-exchange with impurities or free charge carriers even at
low temperatures (5K) lead to a sizeable broadening of the Mu lines which smeares
out the Mu satellite lines \cite{gil_shallow_2001, gil_novel_1999}.
In CdS, the spitting of the inner lines ($\Theta = 109.4^\circ$) is $\Delta\nu_{\rm I} = 0.214(5)$~MHz, 
and the splitting of the outer lines ($\Theta = 0^\circ$) is $\nu_{\rm O} = 0.335(8)$~MHz.
In ZnO, the correponding separations are $\Delta\nu_{\rm I} = 0.413(20)$~MHz, 
and $\Delta\nu_{\rm O} =  0.760(30)$~MHz.

The ratio of ionized to neutral impurities (donors) as a function of temperature $T$  is given by\cite{Seeger}
\begin{equation}\label{eqRatio}
 \frac{N_D^+}{N_D^0} = \frac{N_c}{n}\frac{1}{g_D}\cdot \exp(-\Ei/k_BT)
   \equiv N\cdot \exp(-\Ei/k_BT)
\end{equation}
where $N_c$ is the effective density of states in the conduction band, $n$ is the concentration
of free carriers, $g_D$ is the impurity spin degeneracy, $\Ei$~is 
the ionization energy of the donor, and $N$ is a density-of-states parameter.
We can rewrite in terms of ionized ($f$) and unionized ($1-f$) muonium fractions
\cite{cox_oxide_2006}:
$f/(1-f) = N_D^+/N_D^0$. With this it follows for the neutral Mu$^0$ fraction $f_{\rm Mu}(T)$
\begin{equation}\label{eqTempDependence}
 f_{\rm Mu}(T) = [1-f(T)] = \frac{1}{1+N\cdot \exp(-\Ei/k_BT)}.
\end{equation}
Thus, by measuring the neutral Mu fraction $f_{\rm Mu}(T)$ as a function of temperature
the donor ionization energy $\Ei$ can be determined.
In case of well resolved satellite lines the $\mu$SR asymmetry spectra can be fit by a sum of five Lorentzians
(i.e. exponential relaxation in time domain), with the sum $A_{\rm Mu}$ of asymmetries (amplitudes) of the
four Mu satellite lines, and the asymmetry $A_{\rm D}$ of the so-called diamagnetic signal,
i.e. a $\mu^+$ without
bound electron. The neutral fraction $f_{\rm Mu}(T)$ is then given by
\begin{equation}\label{eqMuFraction}
 f_{\rm Mu}(T) = \frac{A_{\rm Mu}(T)}{A_{\rm Mu}(T) + A_{\rm D}(T)}.
\end{equation}

\section{Results}\label{SecResults}
%
%
%
%
%
%
%
Asymmetry and corresponding frequency spectra for CdS and ZnO are
shown in Figs.~\ref{CdSTimeFFT} and \ref{ZnOTimeFFT}, respectively. The analysis
has been done in the following way.
In practice, it is difficult to derive the temperature dependence of the neutral fraction
by trying to fit five lines to the experimental data over the whole temperature range. 
Problems arise in this procedure in the 
case of poorly resolved satellites or small Mu fractions. Also, with increasing temperature 
spin-exchange processes due to thermally activated charge carriers lead to broadening of the 
Mu satellites and a phase shift of the Mu signal with respect to the
$\mu^+$ signal \cite{senba_electron_2005,gil_spin_2009}. This makes fits in the ionization
regime more difficult: whereas at temperatures $T \lesssim$~15~K the CdS and ZnO data could be fit
with five lines, where we fixed the splitting of the Mu lines to the known values,
this procedure didn't work well in the ionization regime. Therefore, 
we simplified the analysis by an approximation: in CdS the $\mu$SR asymmetry spectra $A(t)$ 
were fit over the whole temperature range by the sum of two exponentially decaying components:
\begin{equation}\label{eqAsymTwoComp}
 A(t) = [A_{\rm D} \exp(-\lambda_{\rm D} t) + A_{\rm Mu} \exp(-\lambda_{\rm Mu} t)]\cos(\omega t + \phi),
\end{equation}
where $\lambda_{\rm D}$ is the relaxation rate of the central, diamagnetic line which we fixed to
the high temperature value (where Mu is ionized, $\lambda_{\rm D} \lesssim 0.01$~$\mu s^{-1}$),
$\omega = \gamma_\mu B$ is the $\mu^+$ precession frequency in the applied magnetic field $B = 10$~mT and
$\gamma_\mu/2\pi = 135.5$~MHz/T is the gyromagnetic ratio of the muon, and $\phi$ is a detectors phase of the 
corresponding decay positron detector. In this way the temperature dependent 
Mu fraction $f_{\rm Mu}(T)$ can be calculated
according to Eq.~\ref{eqMuFraction}, which is then used to determine the ionization energy by fitting 
Eq.~\ref{eqTempDependence} to $f_{\rm Mu}(T)$. In the case of ZnO and Au/ZnO with poorly resolved satellites 
we further simplified the analysis by  using a single exponentially relaxing component:
\begin{equation}\label{eqAsymOneComp}
 A(t) =  A \exp(-\lambda t) \cos(\omega t + \phi).
\end{equation}
The temperature dependence of $\lambda(T)$ can be well approximated by 
Eq.~\ref{eqTempDependence} \cite{vilao_hydrogen_2011}: 
\begin{eqnarray}\label{eqLambdaOneComp}
 \lambda(T) & = & f_{\rm Mu}(T)\lambda_{\rm Mu}(T) + (1-f_{\rm Mu}(T))\lambda_{\rm D}  \nonumber \\
       & \simeq & f_{\rm Mu}(T)\lambda_{\rm Mu}(0) + \lambda_{\rm D},
\end{eqnarray}
and we verified this procedure for CdS by comparing this analysis method with the two-component fits of
Eq.~\ref{eqAsymTwoComp}: both methods yield the same ionization energies within experimental errors.

\subsection{CdS}\label{SecResultsCdS}
\begin{figure}[ht]
\includegraphics[width=0.95\linewidth]{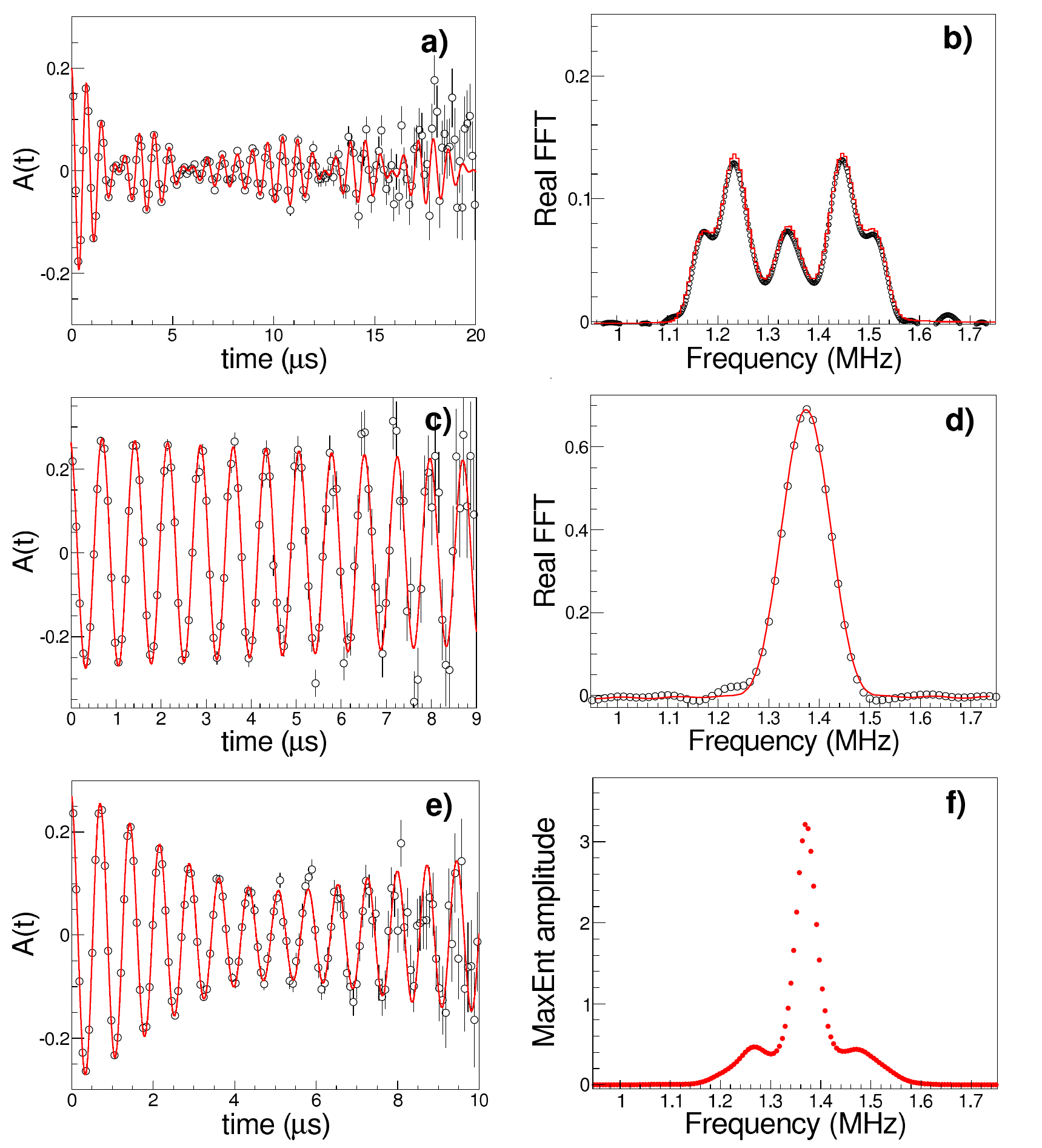}
\caption{CdS (0001), 10 mT applied parallel to a $\langle 0001\rangle$ direction, $\mu$SR asymmetry spectra A(t) 
 and corresponding real part of fast Fourier transform (Real FFT).
 a) and b) 5~K, bulk $\mu$SR (GPS instrument, Muons-On-REquest (MORE) mode \cite{abela_muons_1999}) 
 ($\langle \rm z\rangle\sim$ 280 $\mu$m). c) and d) 5~K, implantation energy 25~keV, virgin polished sample 
 ($\langle \rm z\rangle\sim$ 140~nm). e) and f) 5~K, implantation energy 23 -- 26.5~keV, etched sample.
 The spectrum in f) is obtained by a maximum entropy fit\cite{riseman_maximum_2003I,riseman_maximum_2003II} 
 to the data in four positron counters. The solid red lines are fits to the data as described in the text,
 using the program \tt{musrfit} \cite{suter_musrfit_2012}. 
 }\label{CdSTimeFFT}
\end{figure}
\begin{figure}
 \includegraphics[width=1.0\linewidth]{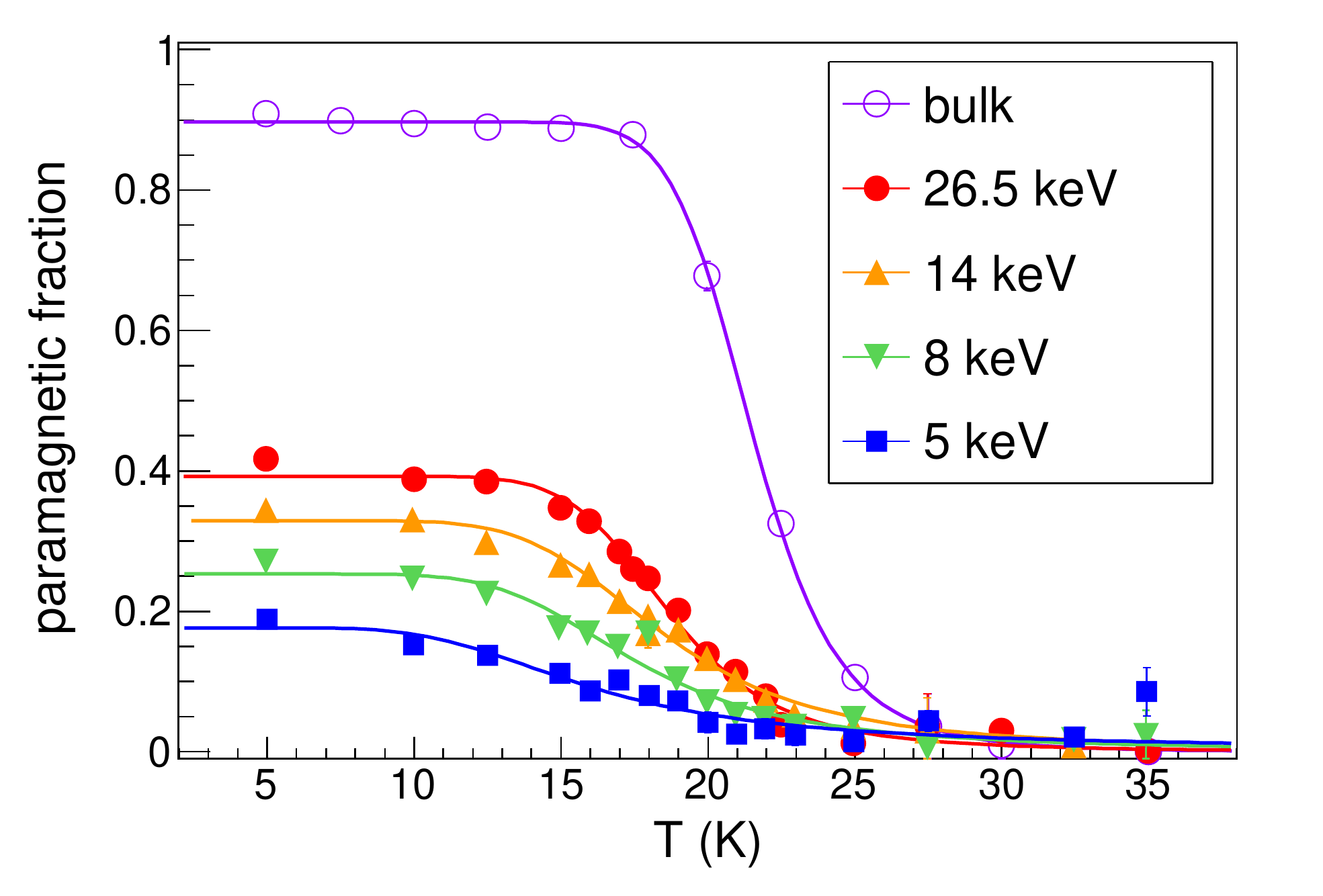}
\caption{CdS (0001), neutral fraction $f_{\rm Mu}(T)$ as a function of temperature $T$ for different implantation
energies. The bulk data are for the virgin sample, and the LE-$\mu$SR data are for the etched sample.
Solid lines are fits of Eq.~\ref{eqTempDependence} to the data to determine the shallow Mu ionization energy. 
 }\label{CdSParaFrac}
\end{figure}
%
Muon spin rotation asymmetry and the corresponding frequency spectra at a temperature of 5~K of the virgin CdS wafer in 
the bulk and at $\langle z \rangle\sim$ 140~nm are shown in Fig.~\ref{CdSTimeFFT}a) -- d). In the bulk a clear
beating is visible reflecting the presence of four shallow Mu satellite lines and the center $\mu^+$ line of
muons which do not form shallow Mu. The two lines with smaller splitting and higher intensity are due to 
shallow Mu at the bonds under 109.4$^\circ$ with respect to the $\langle0001\rangle$ axis, and the two lines
with larger splitting are from shallow Mu at the bond parallel to the $\langle0001\rangle$ axis.
In contrast to the bulk measurements the LEM data do not show any beating which means that in the near-surface
region either shallow Mu does not form, or is strongly suppressed, or -- due to the presence of defects --
fast spin- and/or charge-exchange with a defect electron causes a ``collapse'' of the line splitting resulting 
in a broadening of the diamagnetic line. Recent bulk $\mu$SR experiments in CdS and Si demonstrated the reduction 
of the Mu formation probability in the presence of defects \cite{alberto_electron_2012}.
Rutherford Backscattering Spectrometry (RBS) channeling 
measurements with 2 MeV He nuclei at 
the Institute of Ion Beam Physics at the ETH Zurich revealed that a 
surface layer of at least one $\mu$m depth has a high defect concentration, probably caused by the mechanical
polishing of the CdS wafer. We attribute the absence of the characteristic shallow Mu lines and the slight broadening
of the diamagnetic line in Fig.~\ref{CdSTimeFFT} c) and d) to the presence of these defects.
%
In order to remove the defect-rich surface layer the CdS sample was etched for 55~min at 60$^\circ$~C 
in a 1:1 HCL/H$_2$O solution. The final thickness of the wafer was 0.35(4)~mm, meaning that a total of
$\sim$150~$\mu$m of material was removed. After this procedure the $\mu$SR data show the characteristic 
beating typical for shallow Mu, see Fig.~\ref{CdSTimeFFT} e) and f). Compared to the bulk data the 
shallow Mu fraction is clearly reduced, and the satellite lines are not resolved due to the
shorter time window of 10~$\mu$s in LEM. The reduction of the Mu amplitudes even at highest implantation energies
is likely to be caused by defects which are still present closer to the surface after the etching procedure.

The neutral fraction $f_{\rm Mu}(T)$ as a function of temperature is shown in Fig.~\ref{CdSParaFrac}, where
the $\mu$SR asymmetry spectra were fit with Eq.~\ref{eqAsymTwoComp} to determine the asymmetries
$A_{\rm D}(T)$ and $A_{\rm Mu}(T)$. The decreasing neutral fraction with 
decreasing implantation energy below 26.5~keV is a characteristic normally observed in insulators and
semiconductors \cite{prokscha_formation_2007}. This is attributed to the fact, that a substantial fraction of
Mu is formed by those thermalized $\mu^+$ which may capture one of the excess electrons generated in its own 
ionization track (so-called \textit{delayed} Mu formation). 
The lower the energy the lower the number of track electrons, which reduces the Mu formation probability.
Typically, this \textit{delayed} Mu formation saturates if the stopping depth -- i.e. the track length -- 
of the $\mu^{+}$ is of the order of hundred nanometer \cite{prokscha_formation_2007}. This length scale
fits to earlier observations where the analysis of $\mu$SR experiments with applied electric fields
on bulk insulators suggested a similar length scale for \textit{delayed} Mu formation \cite{eshchenko_excess_2002}.
Bulk $\mu$SR experiments on CdS with an applied electric field
showed that the recombination of a $\mu^+$ 
with a track electron is highly suppressed at relatively weak electric fields of
about 8~kV/cm \cite{eshchenko_excited_2003}.
As we will show below the electric fields due to band bending in CdS at mean implantation
depths $\langle \rm z\rangle <$~40~nm (implantation energy $<$~5~keV) are 6-8~kV/cm. This means that 
the near-surface electric field additionally suppresses Mu formation.

The neutral fraction begins to drop at lower temperature the closer the muons stop to the surface. This
reflects the decrease of the Mu ionization energy on approaching the surface and will be discussed 
in Sec.~\ref{SecResultsIII}.

\subsection{ZnO}\label{SecResultsZnO}
\begin{figure}[ht]
\includegraphics[width=0.95\linewidth]{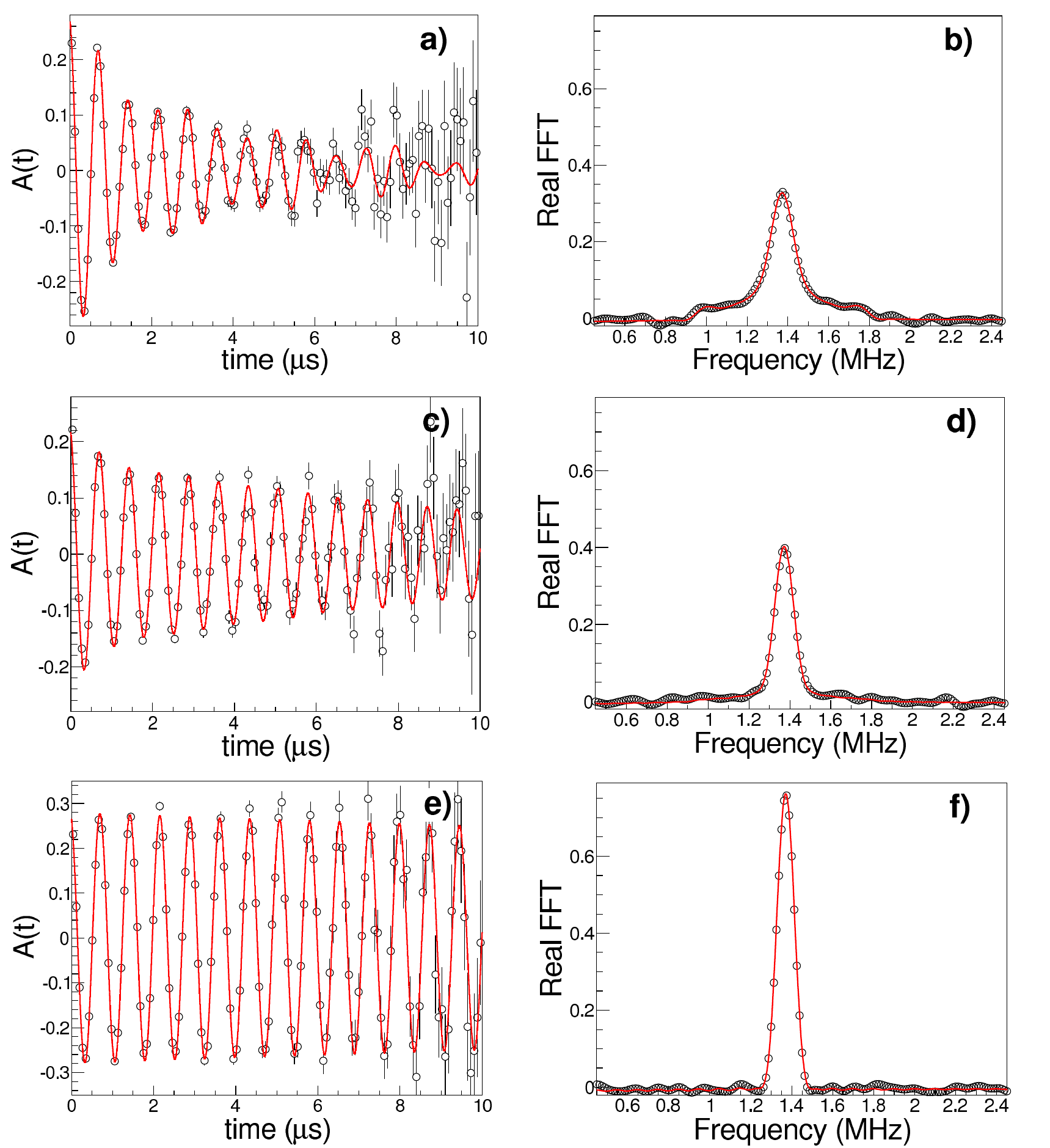}
\caption{ZnO (0001), 10 mT applied parallel to a $\langle 0001\rangle$ direction, 
 $\mu$SR asymmetry spectra A(t) and corresponding real part of fast Fourier transform 
(Real FFT) of one of the positron counters. a) and b) 10~K, implantation energy 17.5~keV 
 ($\langle \rm z\rangle\sim$ 82 nm). c) and d) 10~K, implantation energy 2.5~keV 
 ($\langle \rm z\rangle\sim$ 15 nm). e) and f) 60~K, implantation energy 17.5~keV.
 The solid red lines are fits to the data as described in the text, using the program 
 \tt{musrfit} \cite{suter_musrfit_2012}.
 }\label{ZnOTimeFFT}
\end{figure}
\begin{figure}
\includegraphics[width=1.0\linewidth]{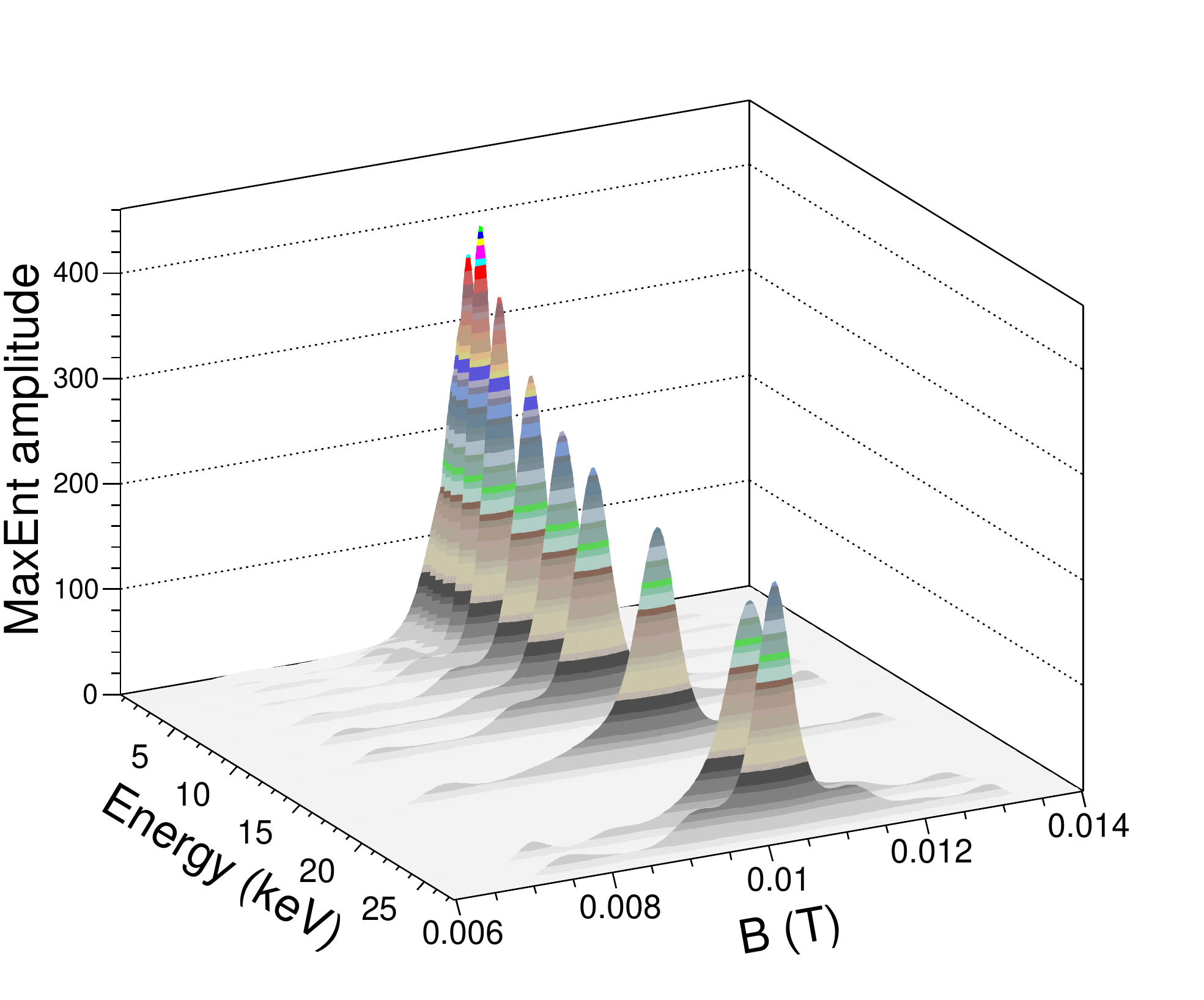}%
\caption{ZnO (0001), 10 K, maximum entropy \cite{riseman_maximum_2003I,riseman_maximum_2003II} 
spectra as a function of implantation energy.
On lowering the implantation energy the diamagnetic peak increases at the expense of the shallow Mu 
satellite peaks.}\label{ZnOMaxEnt}
\end{figure}
\begin{figure}
\includegraphics[width=0.98\linewidth]{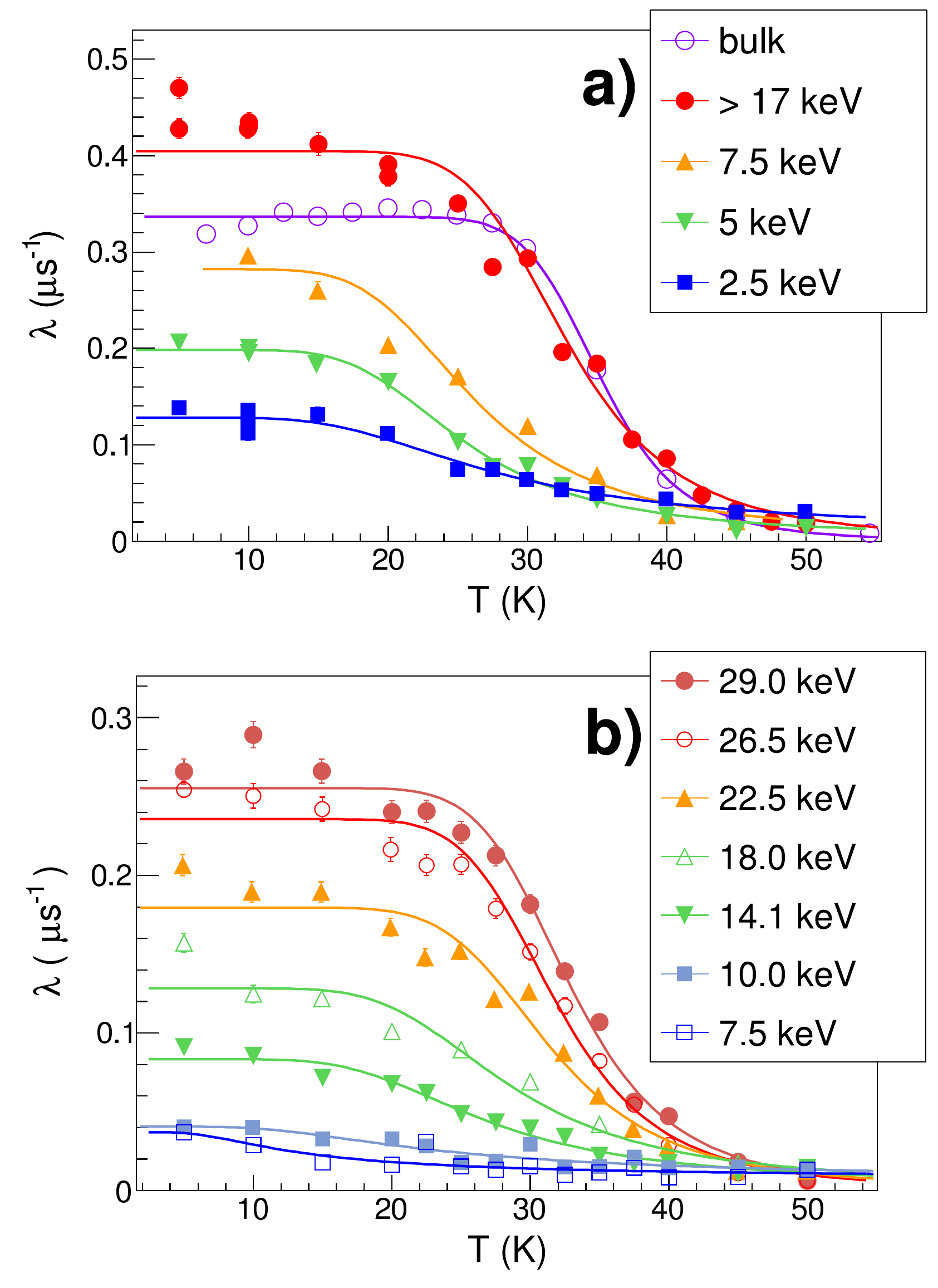}
\caption{a) ZnO (0001), single component exponential depolarization rate $\lambda$ as a function of temperature 
for different implantation energies.
Solid lines are fits of Eq.~\ref{eqLambdaOneComp} to the data to determine the shallow Mu ionization energy.
b) 20 nm Au on top of ZnO (0001). 
}\label{ZnOLambda1expo}
\end{figure}

Compared to the CdS data the shallow Mu lines in ZnO at 10~K are significantly broadened and
unresolved, as shown in Figs.~\ref{ZnOTimeFFT} and \ref{ZnOMaxEnt}. The 10-K data in Figs.~\ref{ZnOTimeFFT}a) and c)
were fit with five frequencies and fixed splitting of the shallow Mu lines, and the 60-K data in
Fig.~\ref{ZnOTimeFFT}e) -- where shallow Mu is ionized -- were fit with a single exponentially relaxing component. 
The bulk data are very similar to the 17.5~keV data of Fig.~\ref{ZnOTimeFFT}a), also revealing unresolved
shallow Mu lines. Figure~\ref{ZnOMaxEnt} shows the frequency spectra at 10~K as a function of implantation
energy, obtained by a maximum entropy fit of the time domain data \cite{riseman_maximum_2003I,riseman_maximum_2003II}.
Similar to the CdS data a clear increase of the diamagnetic line at the expense of the shallow Mu fraction
is visible at decreasing implantation energies. We attribute this as well to the decreasing probability for
\textit{delayed} Mu formation due to the decreasing number of track electrons, and the presence of an electric
field at the surface due to band bending. The higher maximum entropy amplitude of the diamagnetic line
at 27~keV compared to 25~keV has its origin in the slightly more narrow line width of the diamagnetic signal 
at 27~keV (the integral of the line -- which equals the asymmetry $A_{\rm D}$ of the diamagnetic signal -- 
is the same for both energies).

For the determination of the ionization energies at different depths we use Eq.~\ref{eqAsymOneComp} to
fit the $\mu$SR asymmetry spectra, and we plot the relaxation rate $\lambda$ as a function of temperature
and implantation energy. This is shown in Fig.~\ref{ZnOLambda1expo} for ZnO and the Au/ZnO Schottky 
barrier. The absolute value of $\lambda$ is proportional to the Mu fraction $f_{\rm Mu}(T)$ and 
Eq.~\ref{eqLambdaOneComp} has been used to fit the data of Fig.~\ref{ZnOLambda1expo}. 
The relaxation rates in Au/ZnO are generally
smaller compared to ZnO because of muons stopping in the Au layer which contribute to the diamagnetic signal.
Similar to CdS, the relaxation rate -- i.e. the neutral fraction -- begins to drop at lower temperatures
the closer the muons stop at the surface/interface. This is again a manifestation of the decreasing
ionization energy.

\subsection{Depth dependence of the ionization energy of shallow muonium in CdS and ZnO}\label{SecResultsIII}
\begin{figure}
 \includegraphics[width=0.98\linewidth]{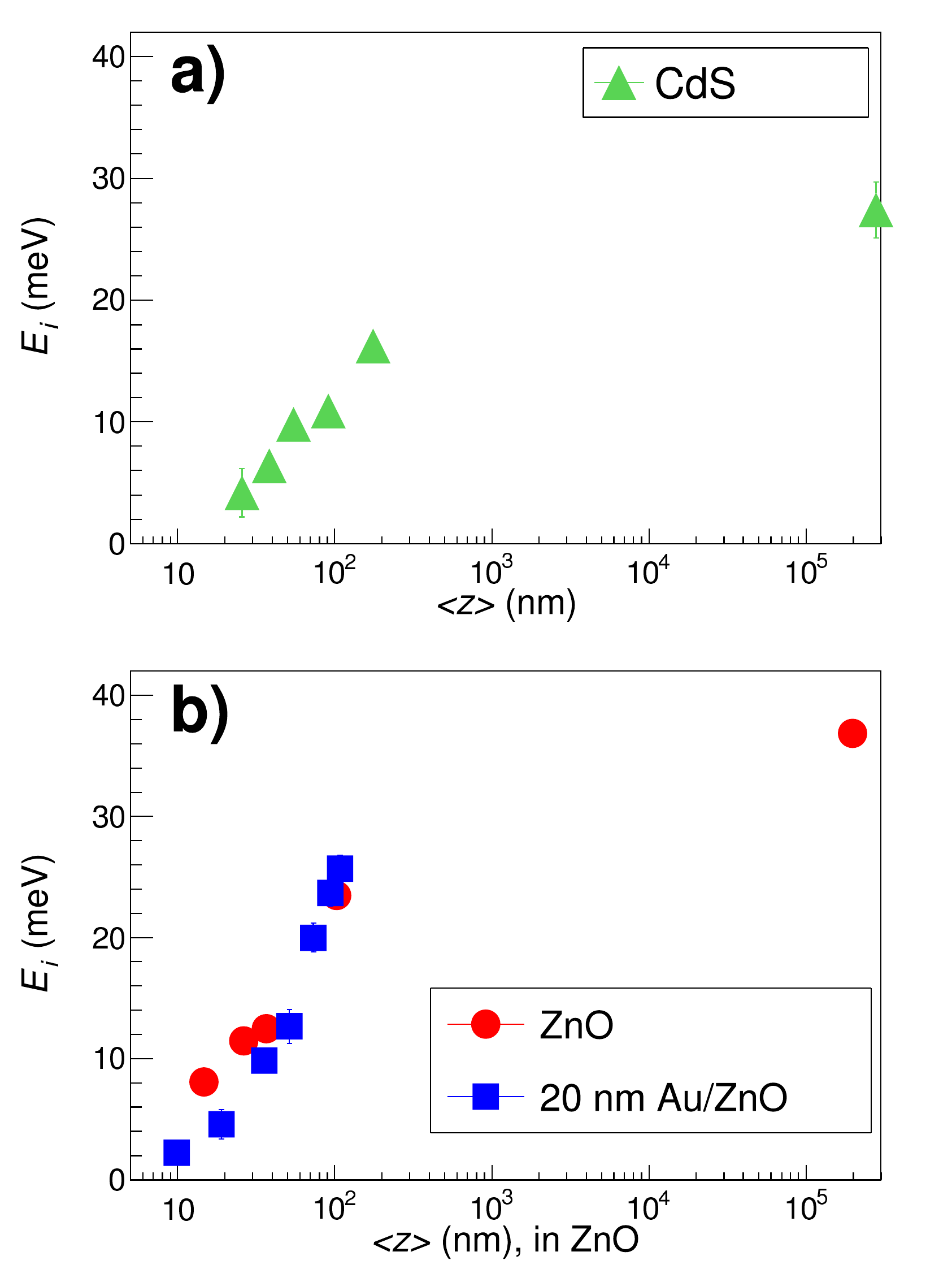}
 \caption{Ionization energy $\Ei$~as a function of mean implantation depth $\langle z\rangle$
 in a) CdS, and b) ZnO and Au/ZnO sample. Note, that mean depths of $\langle z\rangle > 200$~nm
 and $\langle z\rangle \lesssim 100$~$\mu$m are experimentally not accessible due to the lack of
 muon beams with energies between 30~keV and $\sim$~1~MeV. 
}\label{CdSZnOActivationEnergy}
\end{figure}
\begin{figure}
 \includegraphics[width=1.0\linewidth]{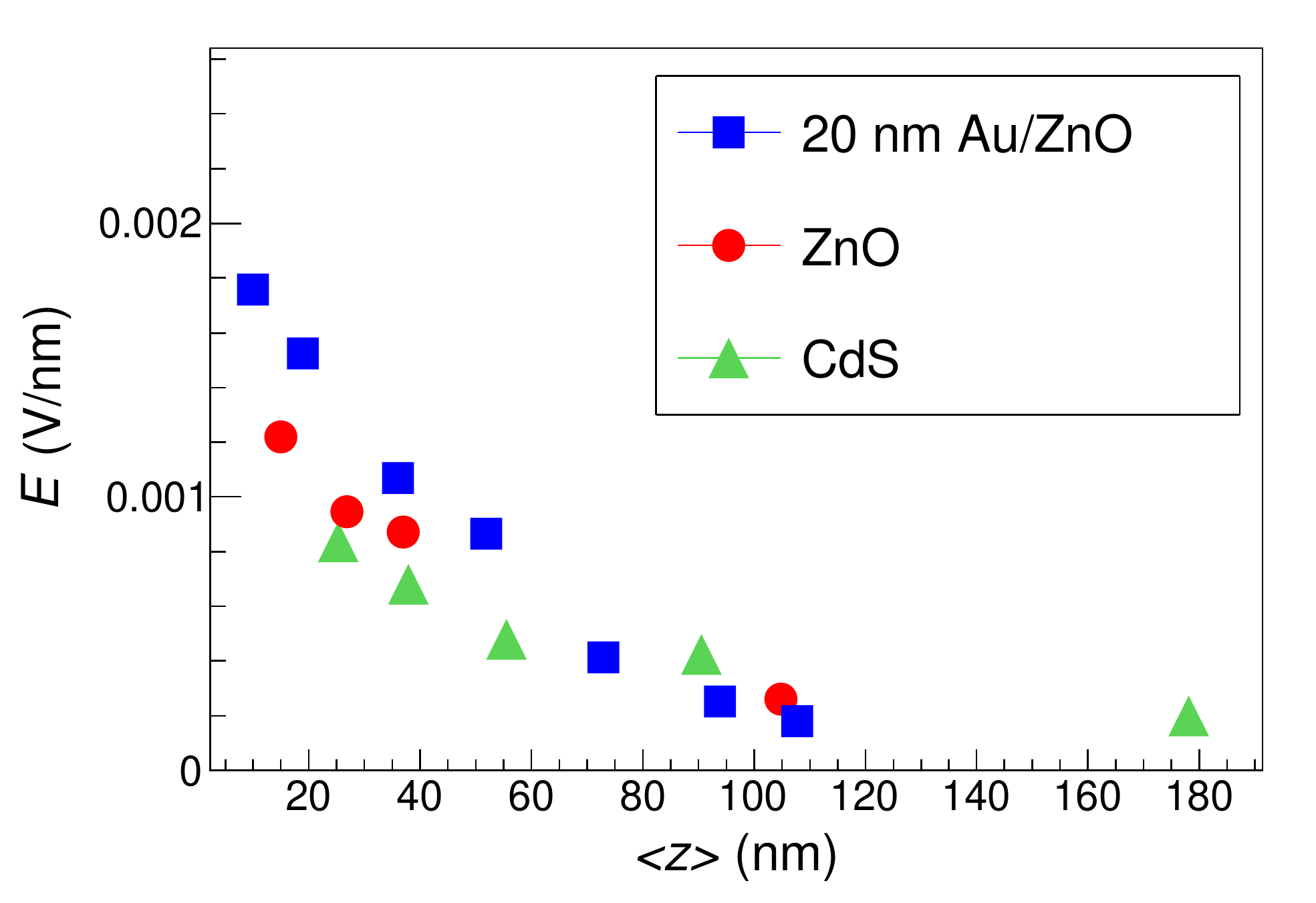}
\caption{Calculated electric field $E$ as a function of mean depth $\langle z\rangle$ 
 in CdS, ZnO, and Au/ZnO.}\label{CdSZnOEField}
\end{figure}
The depth dependence of the ionization energies in CdS and in ZnO, Au/ZnO, are shown in 
Fig.~\ref{CdSZnOActivationEnergy}. Our bulk values ($\langle z\rangle \sim 300$~$\mu$m) are
in agreement with literature data. At the maximum accessible mean depth in LEM of $\sim$~180~nm the
ionization energy is already clearly reduced compared to the bulk value. This reduction
is enhanced on approaching the surface, indicating an increase of the internal electric field.
In Fig.~\ref{CdSZnOActivationEnergy}b) $\langle z\rangle$ denotes the mean depth with respect to
the surface in ZnO, and 
to the metal-semiconductor interface in Au/ZnO.
At the Au/ZnO interface the reduction of $\Ei$ is larger compared to ZnO on a length scale of about
100~nm. This can be attributed to a larger shift of the ZnO electronic bands at the interface
due to the contact to the Au layer, causing a larger band bending, i.e. an enhanced electric field.

The ZnO data suggest a convergence with the bulk ionization energy at a depth of
$\sim 0.5$~$\mu$m, whereas in CdS this length scale appears to be larger ($> 1$~$\mu$m).
The room temperature resistivity of the ZnO wafers is 10~$\Omega$cm, which is hundred times smaller
than the resistivity of the CdS wafer. If we assume that this is caused by a hundred times higher
free charge carrier concentration $n$ in ZnO, the Debye length 
$L_D = \sqrt{\varepsilon_r\varepsilon_0 k_B T/(e^2n)}$ at room temperature -- which is a measure of
the depth of the band bending region -- is expected to be about ten times smaller in ZnO.
The estimated Debye length of $\sim 0.5$~$\mu$m in ZnO at $\sim 30$~K implies a
low temperature charge carrier concentration of $n \sim 5\times 10^{12}$~cm$^{-3}$, which is
consistent with literature data \cite{polyakov_electrical_2006}. The low temperature charge
carrier concentration in CdS is then expected to be in the $10^{10}$~cm$^{-3}$ range to obtain 
a Debye length of the order of $\mu$m.

Using the data of Fig.~\ref{CdSZnOActivationEnergy} and the simple one-dimensional model described
in the appendix the electric field as a function of $\langle z\rangle$ can be calculated and is
shown in Fig.~\ref{CdSZnOEField}. For ZnO and Au/ZnO data $\langle z\rangle$ means again the distance
to the surface (ZnO), or to the Au/ZnO interface. The increase of the electric field close to the
Au/ZnO interface due to enhanced band bending is clearly visible. As described in Sec.~\ref{introduction}
a linear increase of the electric field toward the surface/interface is expected. However, the
data indicate a deviation from linear dependence, with a faster increase of the field the closer the
muons stop to the surface/interface. This could have its origin in the broad stopping distribution
of the muons (see Fig.~\ref{MuonStoppingProfile}): in our simple analysis we effectively determine
an ``averaged'' ionization energy. The larger the implantation energy the larger the range
for the averaging, which may cause the observed deviation from linear dependence.

\section{Discussion}
As we noted in Sec.~\ref{introduction} the binding energy at mean depths $\langle z\rangle > 10$~nm 
is marginally affected
by modifications of the wavefunction at the semiconductor surface, position dependent effective 
masses and dielectric constants, and dielectric confinement. The observed changes on $E_i$ in the
depth range of our low-energy $\mu$SR study (10~nm $ < \langle z \rangle <$ 200~nm)
can be naturally explained by assuming the presence of 
an electric field due to band bending. It is then the Poole-Frenkel effect which causes the reduction
of $E_i$, and this allows us to relate the depth-dependent $E_i$ to the electric field profile.

Our results represent 
the first depth profiling of the ionization energy of a solitary hydrogen-like impurity state over a range
of about 200 nm  by means of a \textit{local probe} technique. In this context \textit{local probe}
means that the probe resides at an interstitial or substitutional side \textit{within} the sample,
where it ``observes'' its local environment on a nanometer scale, such as e.g. $\mu$SR, NMR, $\beta$-NMR,
ESR, PAC, or Moessbauer spectroscopy.
The determination of the depth profile of $E_i$ at semiconductor surfaces or interfaces
by low-energy $\mu$SR
requires the detectability of 
the corresponding muonium states, e.g. semiconductors with doping levels 
$\lesssim 10^{17}$ cm$^{-3}$, and not too high defect concentration. In Sec.~\ref{SecResultsCdS}
we 
showed that a significant fraction of Mu in semiconductors is due to {\em delayed} capture
of an electron from the muon's ionization track, where electrons from the track up to distances of
50~nm - 100~nm can be captured. Assuming that the muon electron capture probability is reduced in the presence 
of defects -- because defects in semiconductors usually act as recombination centers for excess 
carriers \cite{schroder_semiconductor_2006} -- 
a rough estimate for the tolerable defect concentration is given by the requirement
that there are no defects in a volume of $\sim (50$~nm$)^3$ surrounding the stopped muon. This means
that the defect concentration should not exceed $\sim 10^{16}$~cm$^{-3}$.

The determination of the electric-field profile from the depth-dependent change in ionization energy is
an indirect method, but it has the advantage that the sample
can be studied as it is. This is different to a ``surface technique'' such as Kelvin probe microscopy (KPM),
which can directly measure the potential profile at the surface of cleaved samples. It has been frequently 
used in semiconductor studies on pn-junctions, heterostructures, transistors and solar 
cells \cite{melitz_kelvin_2011}, with a spatial resolution in the nanometer range.
Unlike $\mu$SR it cannot provide information about the ionization energy of single impurities.


The presented procedure offers the interesting possibility to study the characteristics of shallow impurities
in the presence of other dopants: the implantation of solitary impurities allows to indirectly
sense the intrinsic charge carrier concentration due its effect on the band bending close to 
the surface or an interface.
The direction of band bending is not accessible here because the changes on the ionization 
energy only depend on the absolute value of the electric field.
Since muon spin rotation is contactless and does not need the application of an electric potential
at the surface it provides direct information about intrinsic properties of the semiconductor.

The simplification of the one dimensional model is well justified if we assume that
the minimum of the ionization potential in one direction is the dominant effect on the
measured change of the ionization energy. The full three dimensional (3D) case is discussed by Martin and
co-workers for deep impurity levels \cite{martin_electric_1981}, where the authors calculate the 
electron emission rate from the impurity state in the 3D case. For example, the electron emission rate 
is diminshed in positive
$z$ direction in the situation skeched in Fig. \ref{FigPotential}. On the other hand the electron emission
rate is increased by phonon-assisted tunneling, and pure quantum mechanical
tunneling. 
For deep levels pure tunneling becomes important only at very high fields ($\sim 10^{7}$ V/cm),
and a significant emission rate enhancement occurs only for  fields $\gg 10^{4}$ V/cm \cite{martin_electric_1981},
which are much larger than the electric fields in our experiment.

\section{Conclusions}
In summary we have shown by means of low-energy $\mu$SR that the ionization energy of single shallow 
hydrogen-like muonium states in CdS, ZnO, and Au/ZnO decreases on approaching the semiconductor
surface or interface. 
Compared to the value measured deep in the bulk at $\langle z\rangle\sim 300$~$\mu$m the ionization
energy is diminshed by $\sim 10$~meV at mean depths of 100 -- 150~nm, and further reduced by up to 
25 -- 30~meV at a depth of 10~nm. This reduction is attributed to the presence of electric 
fields (Poole-Frenkel effect) near the surface/interface due to band bending. Other mechanisms potentially
able to cause a change of the ionization energy (modifications of the wave function, position
dependent effective masses and dielectric constants, dielectric confinement) can be excluded in
the investigated depth range. Using a simple one-dimensional model allowed to determine
the near-surface/interface electric field profile inside the sample. This kind of investigation recently
revealed the presence of a shallow hydrogen donor state in SrTiO$_3$ with decreasing ionization energy
at the surface \cite{salman_direct_2014}. It can be extended to semiconductors or semiconductor heterostructures 
with not too high doping levels ($\lesssim 10^{17}$~cm$^{-3}$) and defect concentrations
($\lesssim 10^{16}$~cm$^{-3}$).

\begin{acknowledgments}
 We gratefully acknowledge the technical support of H.P.~Weber, and the contributions of D.G.~Eshchenko
 in the setup of the LEM facility.
 We thank J.M.~Campos Gil, H.V.~Alberto, and R.C.~Vil\~ao for stimulating discussions.
 The $\mu$SR measurements were performed at the Swiss Muon Source S$\mu$S, Paul Scherrer Institut,
 Villigen, Switzerland.
\end{acknowledgments}

\appendix
\section{Calculation of the electric field}
\begin{figure}
 \includegraphics[width=0.83\linewidth]{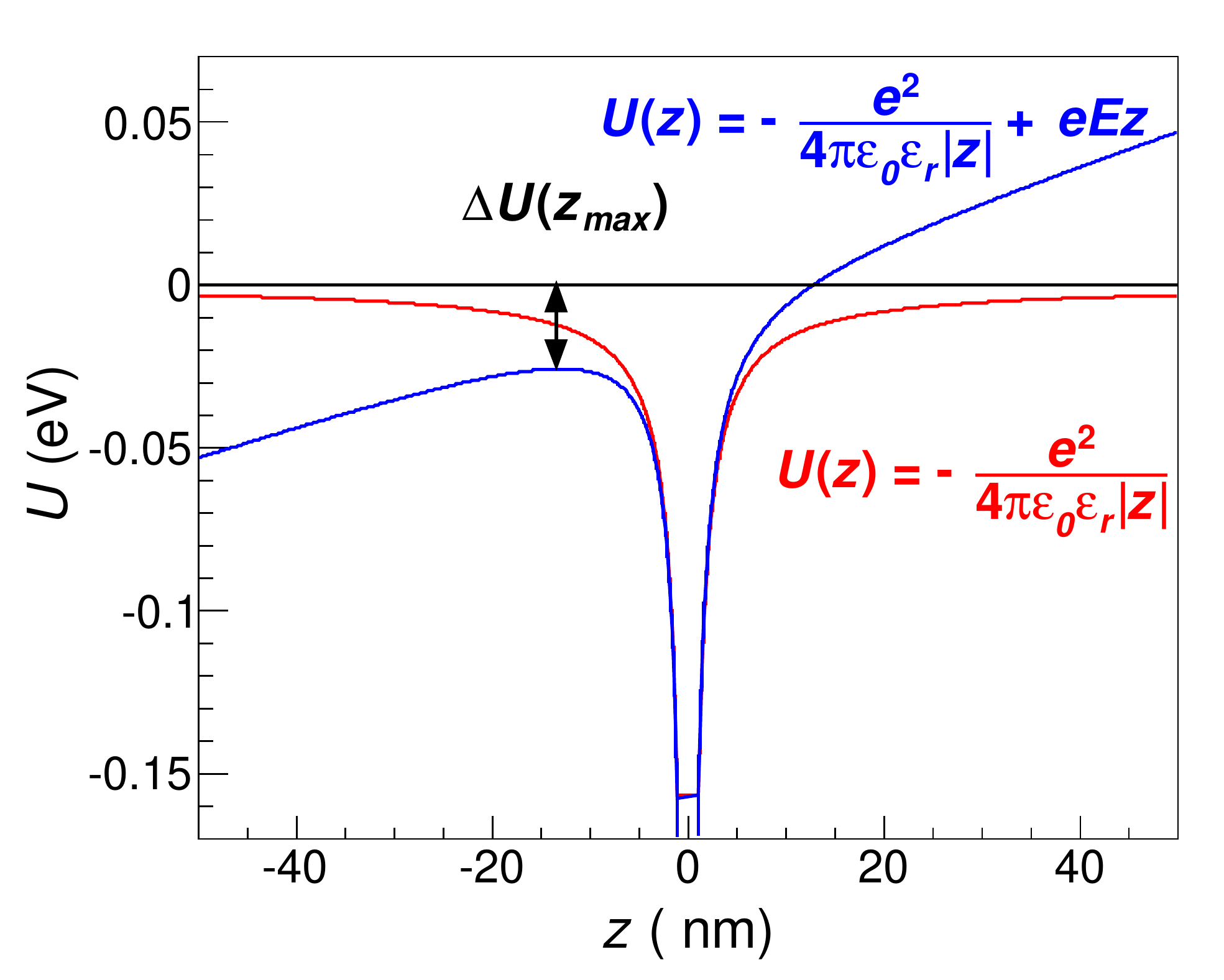}
\caption{Electric potential energy $U(z)$ as a function of $z$ for a point-like charge with and without
 electric field $E$. For the calculation we used the relative permittivitiy 
 of $\varepsilon_r = 8.5$ for ZnO, and $E = 10$~kV/cm = 0.001~V/nm. The double-arrow indicates the
 position of the local maximum $z_{max}$ of $U(z)$ in the presence of an electric field, and the
 reduction $\Delta U(z_{max})$ of the ionization energy.}\label{FigPotential}
\end{figure}

\begin{figure}
 \includegraphics[width=0.83\linewidth]{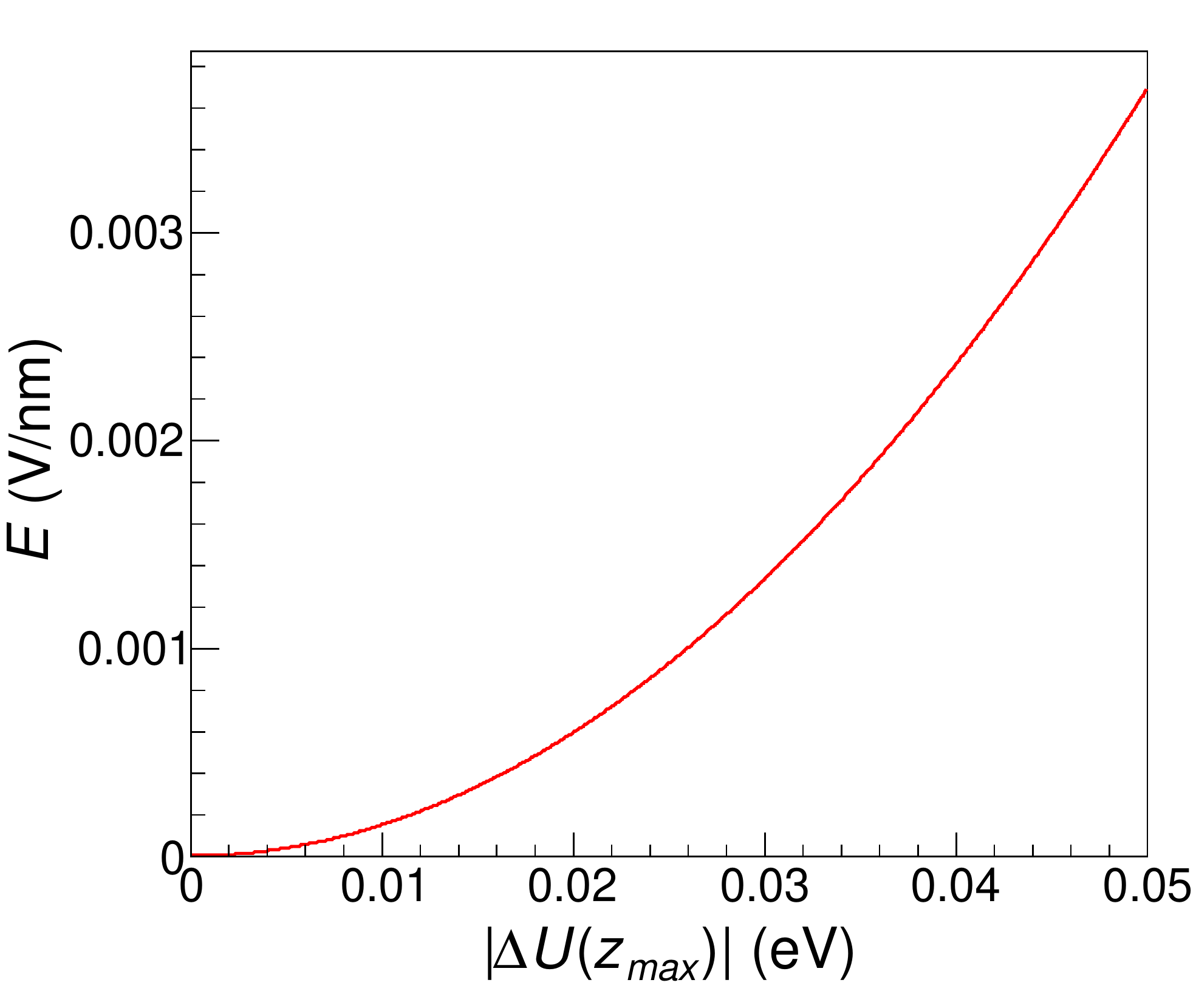}
\caption{Electric field $E$ as a function of $\Delta U(z_{max})$, according to Eq.~\ref{eqA3}, using 
 the relative permittivity of ZnO, $\varepsilon_r = 8.5$.}\label{FigElectricField}
\end{figure}

%
%
%
%
For the calculation of the electric field at the surface we use a one-dimensional approximation since we
are only interested in the maximum reduction of the ionization energy in the presence of an electric field.
The electric potential energy $U$ of a point charge $e$ in the presence of an electric field $E$ along the direction
$z$ can be written as
\begin{equation}\label{eqA1}
  U(z) = -\frac{e^2}{4\pi \varepsilon_0 \varepsilon_r |z|} + eEz,
\end{equation}
where $\varepsilon_0$ is the vacuum permittivity, and $\varepsilon_r$ is the relative permittivity of the
semiconductor.
The change in ionization energy $E_i$ is given by $\Delta U(z_{max})$, as indicated in Fig.~\ref{FigPotential}. 
It can be easily derived from Eq.~\ref{eqA1}\cite{frenkel_pre-breakdown_1938}:
\begin{equation}\label{eqA2}
 \Delta U(z_{max}) = -e\sqrt{\frac{eE}{\pi\varepsilon_0\varepsilon_r}}.
\end{equation}
Solving Eq.~\ref{eqA2} for $E$ we can write the electric field as a function of the reduction of
ionization energy $\Delta U(z_{max})$ (see Fig.~\ref{FigElectricField}):
\begin{equation}\label{eqA3}
 E(\Delta U(z_{max})) = \frac{\pi\varepsilon_0\varepsilon_r [\Delta U(z_{max})]^2}{e^3}.
\end{equation}
We calculated the electric fields shown in Fig.~\ref{CdSZnOEField} with Eq.~\ref{eqA3}, where we
used $\Delta U(z_{max}) = E_i(z>200\mu{\rm m}) - E_i(z)$, and $\varepsilon_r = 8.9$ for CdS, and
$\varepsilon_r = 8.5$ for ZnO.
%

\newpage
\bibliographystyle{prsty}

\pagestyle{empty}

\end{document}